\documentclass[superscriptaddress,prd,aps,showpacs,nofootinbib,showkeywords,eqsecnum,preprint]{revtex4-1}

\usepackage{graphicx,color,amsmath,amsxtra}
\usepackage{epsf}
\usepackage{amssymb}
\usepackage{enumerate}
\usepackage{hhline}
\usepackage{array}
\usepackage{tabularx}
\usepackage[unicode]{hyperref}
\usepackage{graphicx}                
\usepackage{epstopdf}

\begin{document}
\pagestyle{myheadings}

\title{On the Elko inflation model}
\author{Tuan Q. Do }
\email{tuan.doquoc@phenikaa-uni.edu.vn}
\affiliation{Phenikaa Institute for Advanced Study, Phenikaa University, Hanoi 12116, Vietnam}
\begin{abstract}
We argue that it is difficult to investigate the stability of de Sitter inflationary solution in an effective model of Elko fields due to the existence of third-order derivatives of fields, which are unfortunately unconstrained since the number of perturbation equations is not enough. This issue leads to an important question of whether the Elko inflation model is free from an instability or  not.
\end{abstract}
\maketitle
\newpage
\section{Introduction} \label{intro}
Cosmic inflation has played the leading paradigm of modern cosmology in describing the early universe \cite{Starobinsky:1980te,Guth:1980zm,Linde:1981mu}. It is widely believed that an inflationary phase, which is expected to happen in a very short period of time right after the Big Bang, is caused by a scalar field called an inflaton, whose potential should be nearly flat. However, the origin of the inflaton has remained unclear until now. In fact, many approaches for solving this issue have been proposed over several recent decades, e.g., see Refs. \cite{Bassett:2005xm,Mazumdar:2010sa,Nojiri:2010wj,Odintsov:2023weg,Martin:2013tda} for typical examples. 

Very recently, an interesting inflation model based on the so-called Elko fields has been proposed in Ref. \cite{Chen:2024vnl}. It is known that the Elko field is a massive spin-half fermionic field in the $\left(\frac{1}{2},0 \right) \oplus \left(0,\frac{1}{2} \right)$ representation with double Wigner degeneracy, according to the pioneer papers \cite{Ahluwalia:2004ab,Ahluwalia:2004sz,Ahluwalia:2008xi}. See Refs. \cite{Pereira:2014pqa,Ahluwalia:2022ttu} for cosmological implications of the Elko fields.  In the original purpose  \cite{Ahluwalia:2004ab,Ahluwalia:2004sz,Ahluwalia:2008xi}, the Elko fields have been assumed to be associated with the dark matter issue rather than the cosmic inflation one. Surprisingly, the authors of paper \cite{Chen:2024vnl} have demonstrated, possibly with a hint from Ref. \cite{Pereira:2017efk}, that the Elko fields can be an inflaton candidate, provided a suitable field setup and a compatible field redefinition. Very interestingly, it has been shown in Ref. \cite{Chen:2024vnl} that although field equations of these Elko fields are of third-order ordinary differential equations (ODEs) but the Ostrogradsky ghost \cite{Woodard:2015zca} does not appear, ensuring the stability of their model. This point makes the Elko field based inflation model (hereafter we will call it ``the Elko inflation model'' for convenience) very unique and deserves further deep investigations. 

In this paper, we would like to investigate the stability of de Sitter inflationary solution within the model of the Elko fields proposed in Ref. \cite{Chen:2024vnl}. As a result, we will point out that it is impossible to conclude the stability of the de Sitter inflationary solution since the number of perturbation equations is not enough. This result leaves us an important question of the stability of the Elko inflation model. 

This paper will be organized as follows.  (i) Section \ref{intro} has been written for a brief introduction. (ii) Section \ref{sec2} will be used to present a basic setup of Elko inflation model along with its de Sitter inflationary solution in terms of dynamical system. (iii) Section \ref{sec3} will  be devoted to examine the stability of the obtained de Sitter inflationary solution. (iv) Finally, main conclusions will be mentioned in Section \ref{final}. In addition, some calculations will be presented in Appendices \ref{app1} and \ref{app2} for completeness.
\section{Dynamical system of Elko inflation model} \label{sec2}
As said above, Elko inflation has been constructed in a recent paper \cite{Chen:2024vnl}. We refer readers to read this paper for a detailed construction. Here, we would like to begin with  field equations of Elko fields derived in the spatially flat Friedmann-Lemaitre-Robertson-Walker (FLRW) spacetime \cite{Chen:2024vnl},
\begin{align}
\label{field-eq-1}
3H^2 & =M_p^{-2} \left[\frac{1}{2}\ddot\phi +3H \dot\phi + \left(\frac{3}{2}\dot H +\frac{9}{2}H^2 + \frac{d V}{d \phi} \right) \phi +V \right],\\
\label{field-eq-2}
2\dot H +3H^2 & =  M_p^{-2} \left(V- \phi \frac{d V}{d \phi}  \right),\\
\label{field-eq-3}
\dddot\phi +9H\ddot\phi & =  - \left( 9\dot H +27H^2 +4 \frac{d V}{d \phi} +2\phi \frac{d^2 V}{d \phi^2}\right) \dot\phi -  \left(3\ddot H +27H \dot H +27 H^3 +12H \frac{d V}{d \phi}  \right) \phi .
\end{align}
Here $H \equiv \dot a /a$ as the Hubble parameter and $a(t)$ is nothing but a scale factor of FLRW spacetime. It is understood that $\dot a =da/dt$, $\ddot a =d^2 a/dt^2$, and so on.  In addition, $\phi$ is a composite scalar field defined in terms of the Elko field $\lambda$ and its dual field $\overset{\neg}{\lambda}$ such as \cite{Chen:2024vnl}
\begin{equation} \label{field-redefinition}
 \phi \equiv \overset{\neg}{\lambda} \lambda.
 \end{equation}
 It turns out that this field redefinition plays a crucial role in order to make the Elko fields into an inflaton candidate.   It is noted that in order to get the above field equations, the authors of Ref. \cite{Chen:2024vnl} have used the following Einstein-Elko action \cite{Lee:2024sbg}
 \begin{equation} \label{Elko-action}
 S = S_{\rm GR} +S_{\rm Elko} = \int d^4 x \sqrt{-g} \left[ \frac{M_p^2}{2}R -g^{\mu\nu} (\nabla_\mu \overset{\neg}{\lambda}) (\nabla_\nu \lambda )-\frac{1}{4}R \overset{\neg}{\lambda} \lambda -V(\overset{\neg}{\lambda} \lambda)\right],
 \end{equation}
 where $M_p$ is the reduced Planck mass and $V$ is an interacting potential. It is interesting to remark two unusual points in the field equations of the model of Elko fields. The first one is that the Friedmann equation  \eqref{field-eq-1} contains $\ddot\phi$ as well as $\dot H$, while the latter one is about the fact that the equation of motion of $\phi$  \eqref{field-eq-3} involves  $\dddot\phi$. We will come back to these points later when we discuss the stability analysis.

It has been well known that transforming field equations into the corresponding dynamical system will make it more easier to investigate the stability of solutions, e.g., see Refs. \cite{Bahamonde:2017ize,Muller:2017nxg,Kanno:2010nr,Toporensky:2006kc,Barrow:2005qv,Do:2020vdc,Pham:2024fub} for some typical examples. In order to define such a  corresponding dynamical system, we write the scale factor of FLRW metric such as 
\begin{equation}
a(t) \sim \exp [\alpha(t)]
\end{equation}
 and therefore
 \begin{align} 
 H & = \dot\alpha,\\
  \dot H & =\ddot\alpha,\\
   \quad \ddot H & = \dddot\alpha,
 \end{align}
 for convenience. Then, we introduce the following dynamical variables, with hints from Refs. \cite{Muller:2017nxg,Kanno:2010nr,Toporensky:2006kc,Barrow:2005qv,Do:2020vdc,Pham:2024fub}, as follows
\begin{align}
B & = \frac{1}{\dot\alpha^2},\\
 Q & = \frac{\ddot\alpha}{\dot\alpha^2},\\
 \label{def-of-X}
  \quad X & = \frac{\phi}{M_p^2 \dot\alpha^2},\\
   \quad Y& =\frac{\dot\phi}{M_p^2 \dot\alpha^3},\\
    \quad Z& =\frac{\ddot\phi}{M_p^2 \dot\alpha^4},
\end{align}
along with an auxiliary variable defined as \cite{Bahamonde:2017ize}
\begin{equation} \label{equation-U}
U = \frac{\hat\lambda}{\hat\lambda+1}
\end{equation}
with 
\begin{equation} \label{def-hat-lambda}
\hat\lambda = \frac{\phi}{V}\frac{dV}{d\phi}.
\end{equation}
Inversely, $\hat\lambda$ can be defined in terms of $U$ such as
\begin{equation} \label{def-hat-lambda-1}
\hat \lambda = \frac{U}{1-U}.
\end{equation}
It should be stressed that the notation $\hat\lambda$ is completely different from that used to denote the Elko fields, i.e., $\lambda$ and  $\overset{\neg}{\lambda}$.

Autonomous equations are now written in terms of a dynamical time $\alpha \equiv \int \dot\alpha dt$ as follows (see the Appendix \ref{app1} for detailed derivations)
\begin{align}
B' &= -2 B Q,\\
Q' &= \frac{\dddot\alpha}{\dot\alpha^3}-2Q^2,\\
X' & = Y- 2QX,\\
Y'  &= Z - 3QY,\\
Z' &= \frac{\dddot\phi}{M_p^2 \dot\alpha^5} -4QZ,\\
U'  &= M_p^{2} \frac{Y}{B} \frac{dU}{d\phi},
\end{align}
where $' \equiv d/d\alpha$.
Our next goal is to define the remaining terms, $\frac{\dddot\alpha}{\dot\alpha^3}$ and $\frac{\dddot\phi}{M_p^2\dot\alpha^5}$, in terms of the dynamical variables, using the field equations \eqref{field-eq-1}, \eqref{field-eq-2}, and \eqref{field-eq-3}. To do this task, we first combine Eqs. \eqref{field-eq-1} and \eqref{field-eq-2} to have
\begin{equation}
2\dot H+6H^2 = M_p^{-2} \left[ \frac{1}{2}\ddot\phi +3H \dot\phi + \left(\frac{3}{2}\dot H +\frac{9}{2}H^2  \right) \phi +2 V \right],
\end{equation}
which can be written in terms of $\alpha(t)$ as follows
\begin{equation}
2\ddot \alpha +6 \dot\alpha ^2 = M_p^{-2} \left[ \frac{1}{2}\ddot\phi +3\dot\alpha \dot\phi + \left(\frac{3}{2}\ddot\alpha +\frac{9}{2}\dot\alpha^2  \right) \phi +2 V \right].
\end{equation}
Furthermore, multiplying both sides of this equation by $\dot\alpha^{-4}$, we can express it  in terms of  the dynamical variables such as
\begin{equation}
2 B Q + 6 B = \frac{1}{2}Z+ 3Y + \frac{3}{2}QX   + \frac{9}{2} X + 2 B \frac{V}{M_p^2 \dot\alpha^2}.
\end{equation} 

As a result, an useful relation can be figured out from this equation,
\begin{equation}\label{equation-of-V-0}
\frac{V}{M_p^2 \dot\alpha^2} = \frac{1}{B} \left( BQ  -\frac{1}{4}Z -\frac{3}{2}Y -\frac{3}{4}Q X  -\frac{9}{4} X +3B \right).
\end{equation}
Consequently, it turns out that (see Appendix \ref{app1} for detailed derivations)
\begin{align} \label{equation-of-V}
\frac{dV}{d\phi} = \frac{U}{XB\left(1-U \right)}\left( BQ  -\frac{1}{4}Z -\frac{3}{2}Y -\frac{3}{4}Q X  -\frac{9}{4} X +3B \right),
\end{align}
with the help of the definition of $\hat\lambda$ given in Eq. \eqref{def-hat-lambda}.
It is noted that this definition is valid if $X\neq 0$, $B\neq 0$, as well as $U\neq 1$. 
 It is apparent that among the field equations only the equation of motion of scalar field \eqref{field-eq-3} contains the third-order derivative in time of $\alpha$ and $\phi$. In other words, both  $\frac{\dddot\alpha}{\dot\alpha^3}$ and $\frac{\dddot\phi}{M_p^2\dot\alpha^5}$  are constrained only by the following equation (see Appendix \ref{app1} for detailed derivations), 
\begin{align} \label{constraint-equ}
\frac{\dddot\phi}{M_p^2 \dot\alpha^5} + 9 Z = &- \left(9QY +27 Y+ 4 BY \frac{dV}{d\phi} +2M_p^2 XY  \frac{d^2V}{d\phi^2} \right) \nonumber\\
&-\left(3 X \frac{\dddot\alpha }{\dot\alpha^3} +27 QX +27 X +12 BX \frac{dV}{d\phi}  \right).
\end{align}

It now becomes clear  that the absence of $\dddot\alpha$ and $\dddot\phi$ in the Einstein field equations \eqref{field-eq-1} and \eqref{field-eq-2}, which are actually the second-order ODEs, leads to the lack of another constraint equation for $\frac{\dddot\alpha}{\dot\alpha^3}$ and $\frac{\dddot\phi}{M_p^2\dot\alpha^5}$. In principle, we do need two constraint equations for $\frac{\dddot\alpha}{\dot\alpha^3}$ and $\frac{\dddot\phi}{M_p^2\dot\alpha^5}$ in order to figure out them exactly. Despite of this issue, however, it will be shown later that we are still able to figure out fixed point(s) to the dynamical system, even when $\frac{\dddot\alpha}{\dot\alpha^3}$ and $\frac{\dddot\phi}{M_p^2\dot\alpha^5}$ are not determined explicitly.  
 
So far, we have worked out the corresponding dynamical system of the Elko inflation model. Of course, there are some remaining quantities such as $\frac{d^2 V}{d\phi^2}$ and $\frac{dU}{d\phi}$, whose explicit forms will be defined in terms of the dynamical variables once specific forms of $V(\phi)$ are provided. 

Now, we are interested in investigating whether the dynamical system admits an isotropic fixed point, which is indeed equivalent to a de Sitter solution, or not. Mathematically, fixed points of the dynamical system are solutions of the following set of equations \cite{Bahamonde:2017ize,Muller:2017nxg,Kanno:2010nr,Toporensky:2006kc,Barrow:2005qv,Do:2020vdc,Pham:2024fub}
\begin{equation} \label{set-of-equations}
B'=Q'=X'=Y'=Z'=U'=0.
\end{equation}
It turns out that the equation, $B'=0$, implies that 
\begin{equation}
Q=0,
\end{equation}
since $B\neq 0$, which is a condition for the existence of isotropic solution. Hence, the other equations, $Q'=X'=Y'=Z'=0$, imply that
\begin{align}
\frac{\dddot\alpha}{\dot\alpha^3}&=0,\\
Y&=0,\\
Z&=0,\\
\frac{\dddot\phi}{M_p^2 \dot\alpha^5}&=0.
\end{align}

It should be noted that $X\neq 0$, or equivalently $\phi \neq 0$ according to the definition in Eq. \eqref{def-of-X},  has been assumed since we would like to focus on a non-trivial scenario, in which the Elko fields $\lambda$ and  $\overset{\neg}{\lambda}$ do play non-trivial roles, in terms of a non-vanishing $\phi $, in the dynamics of the universe. 

As a result, $X=0$, or equivalently $\phi=0$, will happen to either $\lambda=0$ or $\overset{\neg}{\lambda}=0$, according to the field redefinition \eqref{field-redefinition}.  Consequently, the $\frac{dV}{d\phi}$ shown in Eq. \eqref{equation-of-V} will be undefined. In addition, the action \eqref{Elko-action} will be reduced to that of the Einstein's gravity with a potential, which is now independent of $\phi$ (or  Elko fields). Furthermore, if this potential is chosen to be an effective  constant $\Lambda$, i.e., $V=\Lambda$, then we can obtain the usual de Sitter solution, i.e., $\alpha(t) = \frac{1}{M_p} \sqrt{\frac{\Lambda}{3}}~ t$, which has been well known as the typical cosmological solution of the Einstein's gravity with a cosmological constant,  from the field equations \eqref{field-eq-1} and \eqref{field-eq-2}. However, this de Sitter solution is not driven by the Elko fields and is therefore not our current interest.

Interestingly, the remaining equation, $U'=0$, is automatically satisfied for $Y=0$.   As a result, the constraint equation \eqref{constraint-equ} now reduces to
\begin{equation}
0= 27X +12 BX \frac{dV}{d\phi},
\end{equation}
which yields
\begin{equation}
\frac{1}{B} =-\frac{4}{9} \frac{dV}{d\phi},
\end{equation}
since $X\neq 0$ as well as $B\neq 0$. This equation can be rewritten as
\begin{equation}\label{definition-of-alpha}
\dot\alpha^2 =-\frac{4}{9} \frac{dV}{d\phi}.
\end{equation}
Of course $dV/d\phi$ must be negative since $\dot\alpha^2>0$. 

As a result, the fixed point, $Y=0$, implies that $\dot\phi =0$ and then $\phi = \phi_0= {\text{constant}}$. Hence, $dV\left(\phi_0\right)/d\phi$ will be constant, too. Consequently, taking an integration of this equation will give us a non-trivial de Sitter solution,
\begin{equation} 
\alpha(t) =\frac{2}{3} \sqrt{-\frac{dV\left(\phi_0 \right)}{d\phi} } ~t,
\end{equation}
up to an integration constant. This result is indeed consistent with that obtained in Ref. \cite{Chen:2024vnl}.  Once again, $dV\left(\phi_0\right)/d\phi<0$ is required accordingly.

 It is important to emphasize that the obtained results are really consistent with the general solutions encoded by $H_\pm$ and $\dot H_\pm$ shown in Eqs. (39)-(41) of Ref. \cite{Chen:2024vnl}.  It has been shown that $Q=Y=Z=0$, which correspond to $\ddot\alpha = \dot\phi =\ddot\phi=0$, for the de Sitter fixed point, which we have been interested in. Consequently, Eqs. (39)-(41) of Ref. \cite{Chen:2024vnl} will reduce to
 \begin{align} \label{Hpm}
 H_\pm &= \frac{\pm A }{9\phi -12M_p^2},\\
 A & = \left\{ - \left(9\phi -12M_p^2 \right) \left[ \frac{3}{M_p^2}\left(V \phi -\frac{dV}{d\phi} \phi^2 \right) +4 \left(V + \frac{dV}{d\phi} \phi \right)\right]  \right\}^{1/2},\\
 \label{dotHpm}
 \dot H_+ =\dot H_- & =  \frac{4V}{4M_p^2 -3\phi } -\frac{A^2 }{3 \left( 4M_p^2 -3\phi \right)^2},
 \end{align}
 for this de Sitter fixed point.  Furthermore, $\dot H_\pm $ must be zero for the de Sitter (fixed point) solution since $Q=0$ or  equivalently $\ddot\alpha=0$. As a result, this condition leads to
 \begin{equation}
 12 V \left(4M_p^2 -3\phi  \right) -A^2=0,
 \end{equation}
 according to Eq. \eqref{dotHpm}.
 Thanks to the definition of $A$ shown above, we arrive at
 \begin{equation} \label{positivity}
 3V = \left( 3\phi- 4M_p^2 \right) \frac{dV}{d\phi}.
 \end{equation}
 As claimed in Ref. \cite{Chen:2024vnl}, the positivity of the Hubble parameter $H$ will generically require that 
 \begin{equation} \label{posivity-constraint}
 3\phi -4 M_p^2 <0
 \end{equation}
  or equivalently 
 \begin{equation}
 0<\phi < \frac{4}{3}M_p^2.
 \end{equation} 
 And it turns out that only $H \equiv H_-$ is an accepted solution. 
 Now, we will show that this constraint will lead to the positivity of $V$. Indeed, it appears that  Eq. \eqref{equation-of-V-0} will reduce, for the de Sitter solution, to
 \begin{equation}
 \frac{V}{M_p^2 \dot\alpha^2} = \frac{1}{B} \left( -\frac{9}{4} X +3B \right),
\end{equation}
which can be rewritten as
\begin{equation}
V=\frac{3}{4}\dot\alpha^2 \left(4M_p^2 -3\phi\right) >0.
\end{equation}
Interestingly, the positivity of $V$ will imply, according to Eqs. \eqref{positivity} and \eqref{posivity-constraint}, that
\begin{equation}
\frac{dV}{d\phi} <0,
\end{equation}
which has been addressed above for the existence of the de Sitter solution. To achieve the value of $H\equiv H_-$, we further simplify $A$ as
\begin{equation}
A = \left\{ -\frac{4}{9} \left(9\phi -12M_p^2 \right)^2  \frac{dV}{d\phi} \right\}^{1/2},
\end{equation}
with the help of Eq. \eqref{positivity}.
Consequently, we obtain from Eq.  \eqref{Hpm} that
\begin{equation}
H^2 \equiv H_-^2 = -\frac{4}{9} \frac{dV}{d\phi},
\end{equation}
which coincides with Eq. \eqref{definition-of-alpha}. Since $\dot\phi=0$ as required for the de Sitter (fixed point) solution, we can set $\phi =\phi_0 ={\text{constant}}$, then we have the following results,
\begin{equation}
\frac{dV(\phi_0)}{d\phi} <0, \quad H^2 = -\frac{4}{9} \frac{dV(\phi_0)}{d\phi},
\end{equation}
which are nothing but that derived above.

As a specific demonstration, we will discuss the case considered in Ref. \cite{Chen:2024vnl} 
\begin{equation}
V(\phi)= v_0 +v_1 \phi +v_2\phi ^2.
\end{equation}
It turns out, according to Eq. \eqref{definition-of-alpha}, that
\begin{equation}
\dot\alpha^2 =-\frac{4}{9} \frac{dV(\phi_0)}{d\phi}= -\frac{4}{9} \left( v_1 +2v_2 \phi_0 \right),
\end{equation}
where
\begin{equation}
v_1 +2v_2 \phi_0 <0.
\end{equation}
If we choose $v_0 =M_p^4$, $v_1= -2M_p^2$, and $v_2=1$ as chosen in Ref. \cite{Chen:2024vnl}, then we obtain the following solution,
\begin{align}
\alpha(t) = \frac{2}{3} \sqrt{2\left( M_p^2 - \phi_0 \right)}~ t ,
\end{align}
up to an integration constant, provided that $\phi_0 < M_p^2$.
\section{Stability analysis} \label{sec3}
So far, we have shown that the Elko model can admit a de Sitter solution, provided that $dV\left(\phi_0\right)/d\phi<0$, where $\phi_0$ is a constant. In particular, we have pointed out that this de Sitter solution is nothing but the fixed point solution of the corresponding dynamical system of the Elko model. This result is indeed consistent with that found in the original paper Ref. \cite{Chen:2024vnl}. More importantly, this de Sitter solution can be used to describe an inflationary phase if $|dV\left(\phi_0\right)/d\phi|$ is large enough. 

 In this section, we would like to investigate the stability of the obtained de Sitter inflationary solution. It is worth noting that  the Elko fields have a great chance to be the desired inflaton, if their corresponding de Sitter solution is unstable, according to discussions in Refs. \cite{Pozdeeva:2019agu,Vernov:2021hxo}. In this case, a quasi-de Sitter solution is expected to exist to account the inflationary phase.
 
  First, we perturb the dynamical system around the found de Sitter fixed point as follows (see the Appendix \ref{app2} for detailed derivations)
\begin{align} 
\delta B' &= -2B \delta Q,\\
\delta Q' &= \delta \left( \frac{\dddot\alpha}{\dot\alpha^3}\right) ,\\
\delta X' &= \delta Y -2X \delta Q,\\
\delta Y' &= \delta Z,\\
\delta Z' &= \delta \left(\frac{\dddot\phi}{M_p^2 \dot\alpha^5} \right),\\
\delta U' &= M_p^2 \frac{1}{B} \frac{dU}{d\phi} \delta Y,
\end{align}
where $\delta \left( \frac{\dddot\alpha}{\dot\alpha^3}\right) $ and $ \delta \left(\frac{\dddot\phi}{M_p^2 \dot\alpha^5} \right)$ both obey the only following constraint coming from Eq. \eqref{constraint-equ},
\begin{align} \label{constraint-equ-pertubed}
 \delta \left(\frac{\dddot\phi}{M_p^2 \dot\alpha^5} \right) + 9\delta Z =& -27\delta Y -4B \frac{dV}{d\phi} \delta Y -2M_p^2 X \frac{d^2V}{d\phi^2} \delta Y  - 3X \delta \left( \frac{\dddot\alpha}{\dot\alpha^3}\right) \nonumber\\
 &-27X \delta Q-27\delta X -12 X \frac{dV}{d\phi} \delta B - 12 B \frac{dV}{d\phi} \delta X -12 BX \delta\left( \frac{dV}{d\phi}\right).
\end{align}

Now, a serious issue emerges as follows. It appears that only this constraint equation is not enough to determine explicitly $\delta \left( \frac{\dddot\alpha}{\dot\alpha^3}\right) $ and $ \delta \left(\frac{\dddot\phi}{M_p^2 \dot\alpha^5} \right)$ in terms of other perturbed dynamical variables such as $\delta B$, $\delta X$, and other ones. In other word, we do not have enough number of the perturbation equations to examine the stability of the found fixed point. This indicates that at least one of these two $\delta \left( \frac{\dddot\alpha}{\dot\alpha^3}\right) $ and $ \delta \left(\frac{\dddot\phi}{M_p^2 \dot\alpha^5} \right)$ is not constrained tightly, i.e., can be arbitrary.  Therefore, we will be unclear to determine whether the found fixed point is an attractor point or not. Then, the stability of the de Sitter inflationary solution still remains mystery. 

Before ending this section, it is important to note that we should not take a time derivative for  either Eq. \eqref{field-eq-1} or Eq. \eqref{field-eq-2} to have an additional constraint equation for  $ \delta \left(\frac{\dddot\phi}{M_p^2 \dot\alpha^5} \right)$ or $\delta \left( \frac{\dddot\alpha}{\dot\alpha^3}\right) $, respectively. The reason for this claim is due to the fact that additional mode(s) of perturbations could emerge artificially from the additional third-order derivatives.  Additionally, there is no reason to take a time derivative for only one of these two field equations. Indeed, we can always take a time derivative for both equations \eqref{field-eq-1} and \eqref{field-eq-2}. However, if we do such a thing, we will face another serious issue that the number of perturbation equations containing two perturbed quantities, $\delta \left( \frac{\dddot\alpha}{\dot\alpha^3}\right) $ and $ \delta \left(\frac{\dddot\phi}{M_p^2 \dot\alpha^5} \right)$, is three, which is larger than two. This result implies that it is impossible to figure out clearly $\delta \left( \frac{\dddot\alpha}{\dot\alpha^3}\right) $ and $ \delta \left(\frac{\dddot\phi}{M_p^2 \dot\alpha^5} \right)$ from these three equations. 
\section{CONCLUSIONS} \label{final}
We have pointed out that it is impossible to judge the stability of the de Sitter inflationary solution within the Elko inflation model since the number of perturbation equations is smaller than the number of perturbed fields. This result leaves us an important question of the stability of the Elko inflation model, which should be considered seriously. 
\begin{acknowledgments}
The author would like to thank two anonymous referees very much for their very useful comments and suggestions. The author would also like to thank Drs. Siyi Zhou and Haomin Rao very much for their kind correspondences. Prof. Phung V. Dong is greatly  appreciated for his support. This study is funded by the Vietnam National Foundation for Science and Technology Development (NAFOSTED) under grant number 103.01-2023.50. The author is thankful for a warm hospitality from the IBS Center for Theoretical Physics of the Universe (Cosmology, Gravity and Astroparticle Physics Group) of Korea during his visit, when the paper is being finalized. 
 \end{acknowledgments}
\appendix
\section{Autonomous equations}\label{app1}
In this Appendix section, we would like to show how to derive the autonomous equations of the Elko inflation model. It appears that
\begin{align}
B' & \equiv \frac{dB}{d\alpha} =\frac{1}{\dot\alpha} \frac{dB}{dt} = \frac{1}{\dot\alpha} \frac{d}{dt}\left( \frac{1}{\dot\alpha^2} \right) = -2 \frac{\ddot\alpha}{\dot\alpha^4}= -2 B Q,\\
Q' &\equiv \frac{dQ}{d\alpha} =\frac{1}{\dot\alpha}\frac{dQ}{dt} = \frac{1}{\dot\alpha} \frac{d}{dt}\left(\frac{\ddot\alpha}{\dot\alpha^2} \right)=\frac{\dddot\alpha}{\dot\alpha^3} - 2 \frac{\ddot\alpha^2}{\dot\alpha^4}= \frac{\dddot\alpha}{\dot\alpha^3}-2Q^2,\\
X' &\equiv \frac{dX}{d\alpha}= \frac{1}{\dot\alpha} \frac{dX}{dt} =\frac{1}{\dot\alpha} \frac{d}{dt} \left(\frac{\phi}{M_p^2 \dot\alpha^2} \right) = \frac{\dot\phi}{M_p^2 \dot\alpha^3} -2 \frac{\ddot\alpha \phi }{M_p^2 \dot\alpha^4}= Y- 2QX,\\
Y' &\equiv \frac{dY}{d\alpha}= \frac{1}{\dot\alpha} \frac{dY}{dt}=\frac{1}{\dot\alpha} \frac{d}{dt}\left(\frac{\dot\phi}{M_p^2 \dot\alpha^3} \right) = \frac{\ddot\phi}{M_p^2 \dot\alpha^4} - 3\frac{\ddot\alpha \dot\phi }{M_p^2 \dot\alpha^5} = Z - 3QY,\\
Z' &\equiv \frac{dZ}{d\alpha}= \frac{1}{\dot\alpha} \frac{dZ}{dt}= \frac{1}{\dot\alpha} \frac{d}{dt}\left(\frac{\ddot\phi}{M_p^2 \dot\alpha^4} \right)  =\frac{\dddot\phi}{M_p^2 \dot\alpha^5} -4\frac{\ddot\alpha \ddot\phi }{M_p^2 \dot\alpha^6}  = \frac{\dddot\phi}{M_p^2 \dot\alpha^5} -4QZ,\\
U' &\equiv \frac{dU}{d\alpha}=\frac{1}{\dot\alpha} \frac{dU}{dt}= \frac{\dot\phi}{\dot\alpha} \frac{dU}{d\phi}  = M_p^{2} \frac{Y}{B} \frac{dU}{d\phi}.
\end{align}
Next, we would like to show how to obtain Eqs. \eqref{equation-of-V} and \eqref{constraint-equ}, which will be useful to the autonomous equations defined above. For the first equation, i.e., Eq. \eqref{equation-of-V}, we use the definition of $\hat\lambda$ in Eq. \eqref{def-hat-lambda} to write
\begin{equation}
\frac{dV}{d\phi} = \hat \lambda \frac{ V}{\phi}.
\end{equation}
Furthermore, thanks to Eq. \eqref{def-hat-lambda-1} as well as Eq. \eqref{equation-of-V-0}, we can rewrite the above equation as
\begin{align}
\frac{dV}{d\phi} = &~ \frac{U}{1-U}  \frac{V}{\phi} \nonumber\\
= &~\frac{U}{1-U} \frac{M_p^2 \dot\alpha^2}{\phi}  \frac{1}{B} \left( BQ  -\frac{1}{4}Z -\frac{3}{2}Y -\frac{3}{4}Q X  -\frac{9}{4} X +3B \right) \nonumber\\\
= &~\frac{U}{1-U}\frac{1}{X}\frac{1}{B} \left( BQ  -\frac{1}{4}Z -\frac{3}{2}Y -\frac{3}{4}Q X  -\frac{9}{4} X +3B \right),
\end{align}
which is identical to Eq. \eqref{equation-of-V}.

Next, we rewrite Eq. \eqref{field-eq-3} in terms of  $\alpha(t)$ as
\begin{equation}
\dddot\phi +9\dot\alpha \ddot\phi  =  - \left( 9\ddot\alpha +27\dot\alpha^2 +4 \frac{d V}{d \phi} +2\phi \frac{d^2 V}{d \phi^2}\right) \dot\phi -  \left(3\dddot \alpha +27\dot\alpha \ddot \alpha +27 \dot\alpha^3 +12\dot\alpha \frac{d V}{d \phi}  \right) \phi .
\end{equation}
By multiplying both sides of this equation by $M_p^{-2} \dot\alpha^{-5}$, we can express it in terms of the dynamical variables as follows
\begin{align}
\frac{\dddot\phi}{M_p^2 \dot\alpha^5} + 9 Z = &- \left(9Q +27 + 4 B \frac{dV}{d\phi} +2 M_p^2 X  \frac{d^2V}{d\phi^2} \right) Y \nonumber\\
&-\left(3  \frac{\dddot\alpha }{\dot\alpha^3} +27 Q +27  +12 B \frac{dV}{d\phi}  \right) X,
\end{align}
which is nothing but Eq. \eqref{constraint-equ}.
\section{Perturbation equations} \label{app2}
In this Appendix section, we would like to present simple derivations for the perturbation equations. By perturbing the dynamical variables around the found de Sitter fixed point with $B\neq 0$,  $X\neq 0$, $Q=0$, $Y=0$, $Z=0$, $\dddot\alpha/\dot\alpha^3=0$, and $\dddot\phi/(M_p^2 \dot\alpha^5)=0$ such as
\begin{align}
&B \to B+\delta B,\quad Q \to Q+\delta Q, \quad X \to X+\delta X, \nonumber\\
& Y \to Y+\delta Y, \quad Z \to Z+\delta Z, \quad U \to U+\delta U,
\end{align}
we obtain the perturbation equations from the dynamical system as follows
\begin{align}
\delta B' &=-2 \left(Q \delta B + B \delta Q \right) = -2 B\delta Q,\\
\delta Q' &= \delta \left( \frac{\dddot\alpha }{\dot\alpha^3}\right) -4Q\delta Q = \delta \left( \frac{\dddot\alpha }{\dot\alpha^3}\right) ,\\
\delta X' &=\delta Y -2\left(X \delta Q+ Q \delta X \right)=\delta Y-2 X \delta Q,\\
\delta Y' &=\delta Z-3 \left(Y\delta Q+Q \delta Y \right) =\delta Z,\\
\delta Z' &=\delta \left( \frac{\dddot\phi }{M_p^2 \dot\alpha^5}\right)-4\left( Z \delta Q+Q \delta Z \right)=\delta \left( \frac{\dddot\phi }{M_p^2 \dot\alpha^5}\right),\\
\delta U' &=M_p^2 \left[ \frac{Y}{B} \delta \left(\frac{dU}{d\phi} \right) +\frac{dU}{d\phi} \left(\frac{\delta Y}{B} - \frac{Y}{B^2}\delta B \right) \right] = M_p^2 \frac{1}{B}\frac{dU}{d\phi} \delta Y.
\end{align}
Finally, Eq. \eqref{constraint-equ} is perturbed to be
\begin{align}
\delta \left( \frac{\dddot\phi }{M_p^2 \dot\alpha^5}\right) +9\delta Z =& -\left[ 9\left(Y\delta Q +Q\delta Y \right) +27\delta Y + 4BY \delta \left(\frac{dV}{d\phi} \right) + 4 \frac{dV}{d\phi}  \left(Y\delta B +B\delta Y \right) \right. \nonumber\\
&\left. +2M_p^2 XY\delta \left(\frac{d^2V}{d\phi^2} \right) +2 M_p^2 \frac{d^2V}{d\phi^2}  \left(Y\delta X+X\delta Y \right) \right] \nonumber\\
&-\left[ 3 \frac{\dddot\alpha }{\dot\alpha^3} \delta X +3 X \delta\left(\frac{\dddot\alpha }{\dot\alpha^3} \right) +27 \left( Q\delta X+X\delta Q \right) +27\delta X \right. \nonumber\\
&\left. +12   \frac{dV}{d\phi}   \left(X\delta B +B\delta X \right) +12 BX\delta \left(\frac{dV}{d\phi} \right) \right] \nonumber\\
=& -27\delta Y -4B\frac{dV}{d\phi}  \delta Y -2M_p^2 X\frac{d^2V}{d\phi^2} \delta Y - 3X \delta\left(\frac{\dddot\alpha }{\dot\alpha^3} \right) \nonumber\\
& -27X \delta Q -27\delta X  -12   \frac{dV}{d\phi}   \left(X\delta B +B\delta X \right) -12 BX\delta \left(\frac{dV}{d\phi} \right) ,
\end{align}
which is identical to Eq. \eqref{constraint-equ-pertubed}.

\end{document}